# *Ab Initio* Device-Driven Screening of Sub-1-nm Thickness Oxide Semiconductors for Future CMOS Technology Nodes

[1,2][‡]Linqiang Xu, [3][‡]Yue Hu, [4]Lianqiang Xu, [5]Lin Xu, [1]Qiuhui Li, [6]Aili Wang, [7]Chit Siong Lau, [1,8,9,10,11*]Jing Lu, [2*]Yee Sin Ang

[1] State Key Laboratory of Mesoscopic Physics and Department of Physics, Peking University, Beijing 100871, P. R. China

[2] Science, Mathematics and Technology (SMT) Cluster, Singapore University of Technology and Design (SUTD), Singapore 487372

[3] Department of Physics, University of Hong Kong, Hong Kong, P. R. China.

[4] School of Physics and Electronic Information Engineering, Engineering Research Center of Nanostructure and Functional Materials, Ningxia Normal University, Guyuan 756000, P. R. China

[5] ECE Department, University of California, Santa Barbara, CA

[6] Zhejiang University-University of Illinois at Urbana−Champaign Institute (ZJU-UIUC), Zhejiang University, Haining 310027, P. R. China

[7] Institute of Materials Research and Engineering (IMRE), Agency for Science, Technology and Research (A*STAR), Singapore 138634

[8] Collaborative Innovation Center of Quantum Matter, Beijing 100871, P. R. China

[9] Beijing Key Laboratory for Magnetoelectric Materials and Devices, Beijing 100871, P. R. China

[10] Peking University Yangtze Delta Institute of Optoelectronics, Nantong 226000, P. R. China

[11] Key Laboratory for the Physics and Chemistry of Nanodevices, Peking University, Beijing 100871, P. R. China

*Corresponding Authors. jinglu@pku.edu.cn; yeesin_ang@sutd.edu.sg
[‡]These authors contribute equally to this work.

## Abstract

Ultrathin oxide semiconductors with sub-1-nm thickness are promising building blocks for ultrascaled field-effect transistor (FET) applications due to their resilience against short-channel effects, high air stability, and potential for low-energy device operation. However, the *n*-type dominance of ultrathin oxide FET has hindered their integration into complementary metal-oxide-semiconductor (CMOS) technology, which requires both *n*-and *p*-type devices. Here we develop an *ab initio* device-driven computational screening workflow to identify sub-1-nm thickness oxide semiconductors for sub-5-nm FET applications. We demonstrate that ultrathin $CaO_2$, CaO, and SrO are compatible with *p*-type device operations under both high-performance (HP) and low-power (LP) requirements specified by the International Technology Roadmap of Semiconductors (ITRS), thereby expanding the limited family of *p*-type oxide semiconductors. Notably, CaO and SrO emerge as the *first-of-kind* sub-1-nm thickness oxide semiconductors capable of simultaneously meeting the ITRS HP and LP criteria for both *n*-and *p*-type devices. CaO and SrO FETs outperform many existing low-dimensional semiconductors, exhibiting scalability below 5-nm gate length. Our findings offer a pioneering effort in the *ab initio*, device-driven screening of sub-1-nm thickness oxide semiconductors, significantly broadening the material candidate pool for future CMOS technology nodes.

# 1. Introduction

Transistor scaling is pivotal in driving the information age. Field-effect transistors (FETs) have been aggressively scaled down to the nanometer regime in the past decade to meet the ever-growing demands for device miniaturization and ultracompact integration in complementary metal-oxide-semiconductor (CMOS) technology.[1, 2] While innovative device architectures such as FinFET,[3] gate-all-around (GAA),[4] and complementary FET (CFET)[5] provide promising pathways for future silicon technology nodes, silicon itself struggles at gate lengths below 12 nm (or sub-1-nm *technology nodes*) due to short-channel effect (SCE) and severe carrier mobility degradation at ultrathin channel thickness.[6, 7] As a potential future transistor building block, two-dimensional (2D) semiconductors may offer a route to tame the SCE and mobility degradation bottlenecks of silicon owing to their ultrathin bodies composed of only a few atomic planes and their dangling-bond-free surfaces.[7-11] The potential of 2D semiconductors is recently reinforced by the *International Roadmap for Devices and Systems* (IRDS) 2023,[12] which positions 2D channel FETs as a prospective channel material for future technology nodes beyond 2028.[13] Identifying high-performance 2D semiconductors, particularly those beyond the well-known transition metal dichalcogenides such as $MoS_2$ and $WS_2$, is thus a crucial pursuit for diversifying the candidate materials for developing the future generations of CMOS technology powered by 2D materials.

Ultrathin oxide semiconductors with sub-1-nm thickness have attracted much attention recently for ultrascaled transistor applications as they inherit the advantages of 2D ultrathin body, as well as the environmental stability and energy-efficiency of oxide semiconductors.[14-21] Ultrathin oxides such as $TiO_2$, $TeO_2$ and $In_2O_3$ with thickness down to sub-1 nm have been experimentally shown to exhibit excellent device performance in few-$\mu$m gate length ($L_g$) devices such as high carrier mobility (> 100 $cm^2$ $V^{-1}$ $s^{-1}$), drain current (up to $10^4$ µA/µm), and transconductance (> $10^3$ µS/µm).[17-19, 22] Theoretical studies further suggest the capability of sub-12-nm-$L_g$ *n*-type devices of $TeO_2$,[23, 24] $Ga_2O_3$,[25, 26] and ultrathin $In_2O_3$ (thickness of 0.43 nm)[27] in high-performance (HP) and low-power (LP) electronics as defined by the *International Technology Roadmap for Semiconductors* (ITRS)[28] and the more recent IRDS[12]. However, ultrathin oxides compatible with sub-12-nm-$L_g$ *p*-type FET are scarce, having only been theoretically predicted in bilayer $TeO_2$ recently.[24] Furthermore, ultrathin oxides capable of simultaneously delivering *n*-type and *p*-type FET for both HP and LP applications remain elusive thus far. Such "*np*-compatible" oxide semiconductors can significantly reduce the complexity of CMOS fabrication process as only one type of semiconducting channel material is involved.

In this work, we develop an *ab initio device-driven* computational screening framework in pursuance of ultrathin oxides with sub-1-nm thickness for ultrascaled FET applications. We construct a preliminary pool of ultrathin binary oxides from two sources: (1) *layered* oxides from the Material Cloud Two-Dimensional Crystals Database (MC2D)[29, 30] and (2) *nonlayered* oxides derived from 3D nonlayered bulks based on a recent high-throughput screening approach[31]. By combining density functional theory (DFT) simulations, device scale length theory[32], and quantum transport device simulations[33], we identify 3 candidates, namely $CaO_2$, CaO, and SrO, that fulfill both HP and LP requirements under *p*-type device configurations, thus expanding the rather scarce pool of ultrathin oxides for *p*-type FET application. Notably, we show that CaO and SrO are the *first-of-kind* sub-1-nm thickness oxide semiconductors that simultaneously meet both ITRS HP and LP requirements under both *n*-type and *p*-type device configurations. The CaO and SrO device performance surpasses many low-dimensional semiconductor and advanced silicon-based technology node transistors, and can be further downscaled below 5-nm $L_g$. These findings reveal the potential of sub-1-nm thickness oxide semiconductors for future CMOS technology nodes and shall open a new frontier in designing ultrascaled transistor based on ultrathin oxide semiconductors.

## 2. Results

**2.1 Preliminary Screening of Sub-1-nm Thickness Oxides**

Our screening workflow is shown in **Figure 1a**. The binary ultrathin oxides are compiled from two sources: (1) layered 2D oxides from the MC2D database with known 3D bulk layered parents;[29, 30] and (2) recently reported nonlayered monolayer oxides derived from 3D bulk nonlayered parents.[31] Such nonlayered materials can potentially be fabricated via CMOS-compatible techniques such as atomic layer deposition for ultrathin $In_2O_3$ and sputtering for ultrathin indium tin oxide (ITO).[15-17, 20] Previous study has employed DFT combined with non-equilibrium Green's function (NEGF) to characterize 100 most promising 2D layered semiconductors (note that only 4 layered binary oxides, i.e. $NiO_2$, $PtO_2$, $Pb_2O_2$, and $TlO_2$), focusing on assessing the device transport characteristics such as on-state current ($I_{on}$) and the subthreshold swing (SS).[34] Here we expanded the device characterizations to more holistically cover other key performance indicators of FET such as delay time ($\tau$) and power delay product (PDP) as well as to understand the gate length scalability below 5 nm of sub-1-nm-thick oxide semiconductors.

**Layered oxides.** Starting from 3172 species of layered 2D materials in MC2D, we select binary oxides with the number of atoms per unit cell smaller than 10 for better ease of fabrication and device simulations, which substantially reduces the candidate pool to 33 species. Previous study suggest that semiconductor with a bandgap larger than 0.4 eV is required for achieving efficient on-off switching.[35] We thus set a PBE bandgap criteria ranging from 0.4 eV to 5 eV, which yields 13 candidates (the feasibility of PBE functional in accessing bandgap is discussed in the **Method** section). Those with imaginary phonons, as calculated in the MC2D database, are removed due to the dynamical instabilities. Candidates that contain toxic elements (i.e. As, Pb, Hg, and Tl) are further excluded. Such a screening process leads to a final candidate pool of 7 layered ultrathin oxides (see **Table S1**).

**Nonlayered oxides.** The recently computationally discovered nonlayered ultrathin oxides are also included.[31] High-throughput computational screening revealed the fabrication possibility of a large family of ultrathin *nonlayered* oxides which can potentially be obtained from their 3D *nonlayered* bulk parents. Such nonlayered oxides greatly expand the candidate pool for our screening process. We start with 41 dynamically and thermally stable nonlayered oxide semiconductors classified as "easily exfoliable" and "potentially exfoliable" oxides (i.e. binding energy smaller than 3 $J/m^2$). One ferromagnetic and six antiferromagnetic oxides are then excluded to avoid the complexity of magnetism. Since only the HSE band gaps of the oxides are reported in the original study, we recalculate the PBE bandgaps of the 34 oxides (see **Table S1**). Using the same PBE bandgap criteria of 0.4 eV to 5.0 eV, we obtain a pool of 27 nonlayered oxide candidates. We further note that the nonlayered ultrathin oxide derived from both MgO(211) and MgO(110) are identical. We thus only select MgO(211) as a representative candidate. Similarly, CaO(112) is selected as the representative of CaO(112), CaO(110), CaO(10$\bar{1}$) since the exfoliation of all three crystal planes leads to the same monolayer oxides. Thus, the final nonlayered ultrathin oxide pool is composed of 24 candidates.

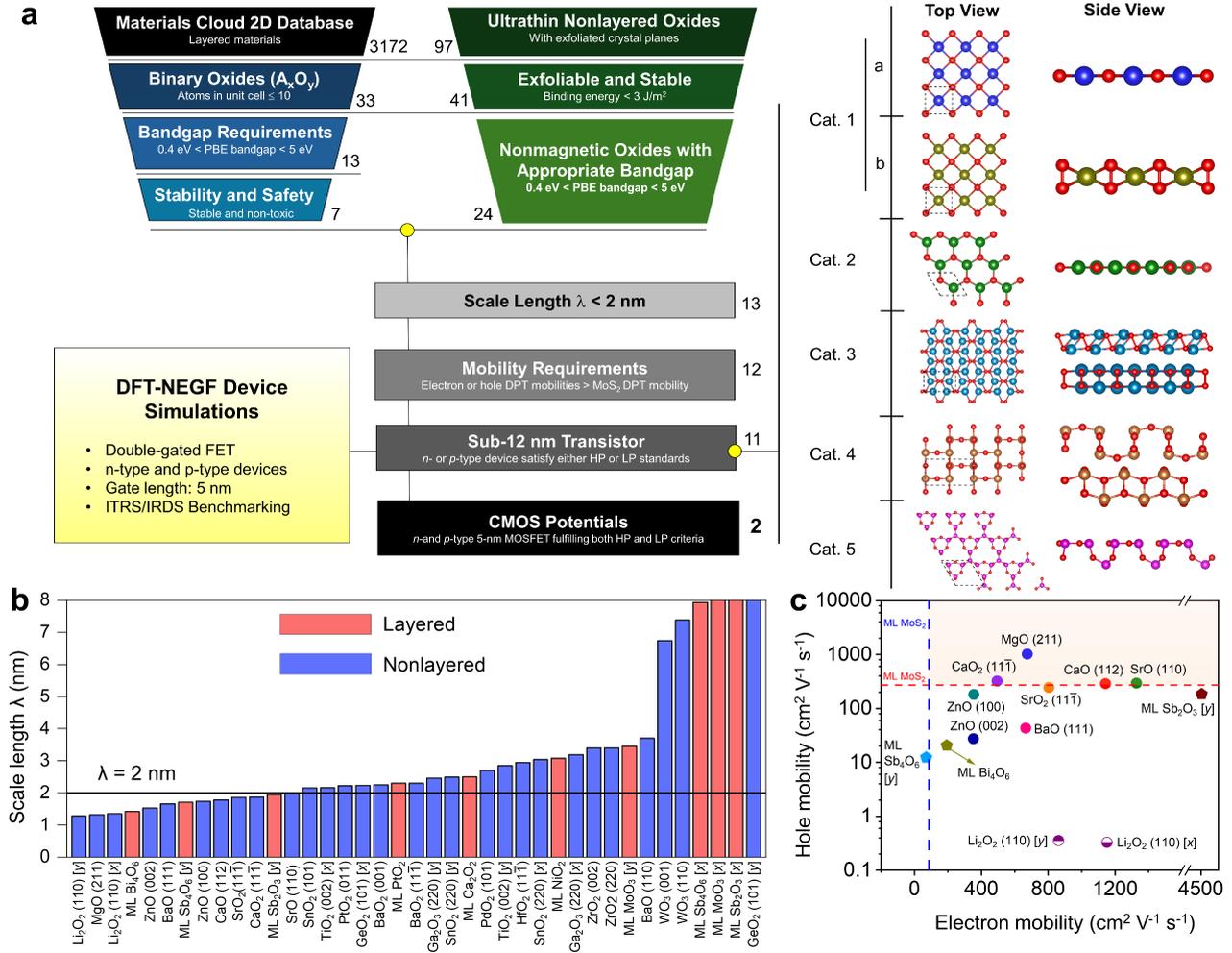

**Figure 1. Workflow of device-driven *ab initio* screening of ultrathin oxide semiconductor.** (a) Screening workflow for ultrathin oxides (left panel). The right panel shows the top and side views of those oxide candidates that satisfy ITRS HP or LP requirements. They are divided into 5 categories by the space group: (Cat. 1a) contains CaO(112), SrO(110), MgO(211), ZnO(100); (Cat. 1b) contains CaO$_2$(11$\bar{1}$); (Cat. 2) contains BaO(111) and ZnO(002); (Cat. 3) contains Li$_2$O$_2$(110) in both *x* and *y* directions; (Cat. 4) contains Sb$_2$O$_3$; and (Cat. 5) contains Bi$_4$O$_6$. The red spheres in the lattice structures represent oxygen atoms, while the other colored spheres denote metal atoms. The black dashed squares indicate the unit cells. (b) Scale length λ of the preliminary filtered oxide candidates. The black line represents the smallest λ value for advanced Si-based FET. (c) Electron and hole mobilities of the filtered oxide candidates by λ. The blue and red lines indicate the calculated electron and hole mobilities of ML MoS$_2$, respectively. For (b) and (c), the anisotropic structures are separated into *x* and *y* directions.

## 2.2 Scale length and DPT mobility

Combining the layered and nonlayered ultrathin oxides, we obtain a total of 31 candidates. To further narrow down the candidates, the intrinsic material properties related to the device performance are computed. The first parameter is the scale length λ, which depicts the penetrating distance of the electric field generated by the electrodes into the semiconducting channel.[6, 33] A smaller λ thus corresponds to a weaker SCE. For 3D semiconductor FET, λ is defined as $\lambda_{3D} = \sqrt{\frac{t_{ch} t_{ox} \varepsilon_{ch}}{\varepsilon_{ox}}}$, where $t_{ch}$ ($t_{ox}$), and $\varepsilon_{ch}$ ($\varepsilon_{ox}$) are the channel (gate oxide) thickness, and dielectric constant of the channel (gate oxide), respectively. The 3D scale length, however, overestimates the SCE for low-dimensional devices.[32] The scale length theory has been extended to low-dimensional FETs which can be written as:[32]

$$\lambda_{2D} = \frac{\pi \left[ t_{ch} \frac{\varepsilon_{ch}}{\varepsilon_{ox}} + 2 t_{ox} \right]}{A}$$

(1)

where *A* is a geometry-dependent coefficient with a value of 3.13 for 2D dual-gated FET. Since the IRDS criteria are developed for FinFET, GAA FET, and CFET, we use the ITRS 2013 version (denoted as "ITRS") as the benchmarking standard. According to the ITRS criteria on the 2028 horizon, the values of $t_{ox}$ and $\varepsilon_{ox}$ are 0.41 nm and 3.9, respectively, for both HP and LP devices. We calculate the $\lambda_{2D}$ for the 34 candidates based on the calculated $t_{ch}$ and $\varepsilon_{ch}$ (see **Table S2**). **Figure 1b** shows the calculated $\lambda_{2D}$ for both layered (red) and nonlayered (blue) ultrathin oxides. Generally, $L_g$ is expected to be at least 6 times of $\lambda_{3D}$ so to avoid SCE dominating the device performance for silicon FET.[6] The smallest $L_g$ of silicon FETs at sub-1 nm *technology node* is 12 nm, which corresponds to $\lambda_{3D} \sim 2$ nm. We thus impose the same limit of $\lambda_{2D} < 2$ nm as a screening criterion, which leads to a smaller pool of only 13 ultrathin oxides, including the anisotropic $Li_2O_2$ (110) where both its *x* and *y* directions are considered as two separate potential candidates.

Carrier mobility *μ* represents another useful parameter that critically influences the device performance. For oxide semiconductors, low carrier mobility (*μ* typically around 10 cm$^2$ V$^{-1}$ s$^{-1}$) is one of the major obstacles hindering their applications in electronic devices.[14, 22] Here we calculate the electron ($\mu^{(e)}$) and hole ($\mu^{(h)}$) carrier mobility by the deformation potential theory (DPT) method (see **Table S3**). Since the DPT method only considers the scattering from the longitudinal acoustic phonon, the values calculated here should be regarded as the upper limits of the intrinsic carrier mobility. The electron and hole carrier mobility of monolayer (ML) MoS$_2$ (i.e. $\mu^{(e)}_{MoS_2}$ and $\mu^{(h)}_{MoS_2}$, respectively) calculated using the same approach (see **Table S4**) are used as a benchmark for assessing the mobility of the candidates. All candidates with carrier mobility *lower* than that of ML MoS$_2$ are eliminated (see **Figure 1c** where the ML MoS$_2$ mobilities are indicated by the dashed lines), yielding 12 candidates with superior electron mobility to ML MoS$_2$ (i.e. 2 layered and 10 nonlayered oxides), but only 4 nonlayered ultrathin oxides possess higher hole mobility than ML MoS$_2$. It should be noted that none of the layered oxides survive at this stage in terms of hole mobility, which is consistent with the commonly observed scarcity of *p*-type oxide semiconductors with high hole mobility.[16, 19, 36]

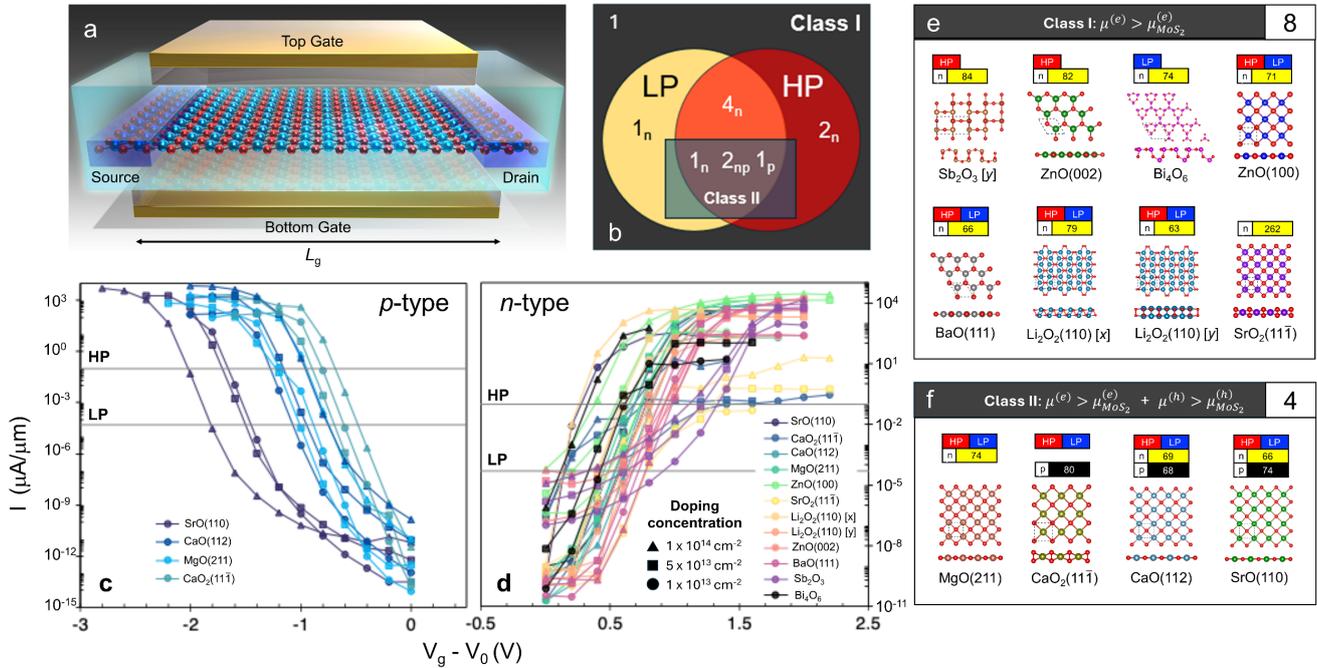

**Figure 2. Sub-1-nm thickness oxide semiconductor FET and current-voltage characteristics.** (a) Schematic diagram of the dual-gated device configuration. (b) Device classifications of *Class I* candidates which have higher electron mobility than MoS$_2$ and of *Class II* candidates which have both higher electron and mole mobilities than MoS$_2$. The subscript of the number of candidates denotes the device type (i.e. n-type or p-type). Transfer curves of the (c) *p*-type and (d) *n*-type transistors at 5-nm $L_g$. The grey horizontal lines indicate the ITRS criteria for HP and LP off-state current $I_{off}$. The transfer curves are shifted by V$_0$ so that they all start from 0 V for visual clarity (see **Supplementary Figure S1** for the original transfer curves without V$_0$ shifting). (e) and (f) show the structures and specify device performance of *Class I* and *Class II* sub-1-nm thickness oxides. The "HP" and "LP" markers denote candidates that fulfill the ITRS HP and LP requirements, respectively. The "*n*" and "*p*" markers denote the compatibility of the candidates with *n*-type and *p*-type device configurations, respectively. The numbers following "*n*" or "*p*" indicate the minimum SS achievable among three different doping concentrations.

## 2.3 Ab initio quantum transport device simulations: ITRS and IRDS benchmarking of $I_{on}$

We perform *ab* initio quantum transport simulations based on a dual-gated device configuration (see **Figure 2a**). The *n*-type and *p*-type transfer curves at $L_g$ of 5 nm are simulated for the 12 candidates with higher electron or hole mobilities than ML MoS$_2$ (**Figures 2c** and **2d** for the transfer curves, and **Figures 2b, 2d,** and **2e** for a summary of the device performance). The device types (*n*-type or *p*-type) are set by controlling the doping carriers at the source and drain electrodes (see **Methods** section). Three different levels of doping concentrations ($N_d$) are used in the device simulations for each candidate i.e. $1 \times 10^{14}$, $5 \times 10^{13}$, and $1 \times 10^{13}$ cm$^{-2}$. According to the ITRS device requirement on the 2028 horizon, the HP off-state current ($I_{off}$), LP $I_{off}$, and supply voltage ($V_{dd}$) are set as 0.1 µA/µm, $5 \times 10^{-5}$ µA/µm, and 0.64 V, respectively. Since the on-state voltage ($V_g^{on}$) can be calculated by $V_g^{on} = V_g^{off} \pm V_{dd}$ ("+" for $n$-type and "-" for $p$-type, $V_g^{off}$ is the off-state voltage), the on-state current ($I_{on}$) can thus be extracted from the transfer curves at $V_g^{on}$.

**Figure 3a** shows the extracted HP *vs* LP $I_{on}$ of the *n*-type ultrathin oxide FETs for the *n*-type transistors at different $N_d$. We identify 1 (1) layered oxide and 8 (7) nonlayered oxides that are suitable for the HP (LP) application. Among those candidates, 7 candidates (i.e. CaO(112), SrO(110), BaO(111), MgO(211), ZnO(100), Li$_2$O$_2$(110)[*x*], Li$_2$O$_2$(110)[*y*]) can simultaneously meet the requirements of both HP and LP under *n*-type device setup. While for the *p*-type device, only three candidates [i.e. CaO (112), CaO$_2$(11$\bar{1}$), and SrO(110)] whose *p*-type device operation fulfills both the HP and LP requirements of ITRS are also identified (**Figure 3b**). Importantly, the discovery of such *p*-type oxide candidates further expands the rather small pool of existing *p*-type oxide semiconductors (such as bilayer TeO$_2$).[19, 24] After comparison, $N_d$ of $5 \times 10^{13}$ cm$^{-2}$ is identified as the optimal concentrations of both *n*-type and *p*-type devices for subsequent discussion and simulation.

In **Figures 3c** to **3f**, we focus on benchmarking the performance of the 4 candidates with superior electron and hole mobilities than ML MoS$_2$ (i.e. CaO$_2$(11$\bar{1}$), MgO(211), SrO(110), and CaO(112) as listed in **Figure 2f**). CaO(112) and SrO(110) (denoted as CaO and SrO in the following for notational simplicity) as sub-1-nm thickness oxide semiconductors that can simultaneously fulfill the HP and LP requirements of $I_{on}$ under both *n*-type and *p*-type device setups (**Figures 3c** and **3d**). The $I_{on}$ of CaO and SrO FETs exhibit better HP and LP performance than multiple low-dimensional semiconductors, such as ML MoS$_2$, ML MoTe$_2$, ML silicane, bilayer TeO$_2$ (both *x* and *y* directions), and Si nanowire (NW) at the same $L_g$ (5 nm).[24, 37-40]

We further benchmark the device performances of the CaO and SrO with that of the advanced Si technology nodes (**Table 1**). The latest IRDS 2023 version has listed the requirements for HP and high-density (HD) devices from 3 to 0.5 nm *technology nodes*.[12] Since the $I_{off}$ for HP and HD transistors is set as 0.01 and $1 \times 10^{-4}$ µA/µm by IRDS, respectively, we reprocessed the data and obtained new $I_{on}$ (**Table 1**). Although there are differences in the structure architectures between our device and advanced Si FETs, such comparison could also be instructive to gauge the oxide device performance in relevance to the IRDS technology node requirements. Under the IRDS criteria, HP and HD $I_{on}$ of both *n*-type and *p*-type CaO and SrO FETs continue to outperform all the Si technology node devices, thus indicating the potential of these sub-1-nm thickness oxides for advanced CMOS technology.

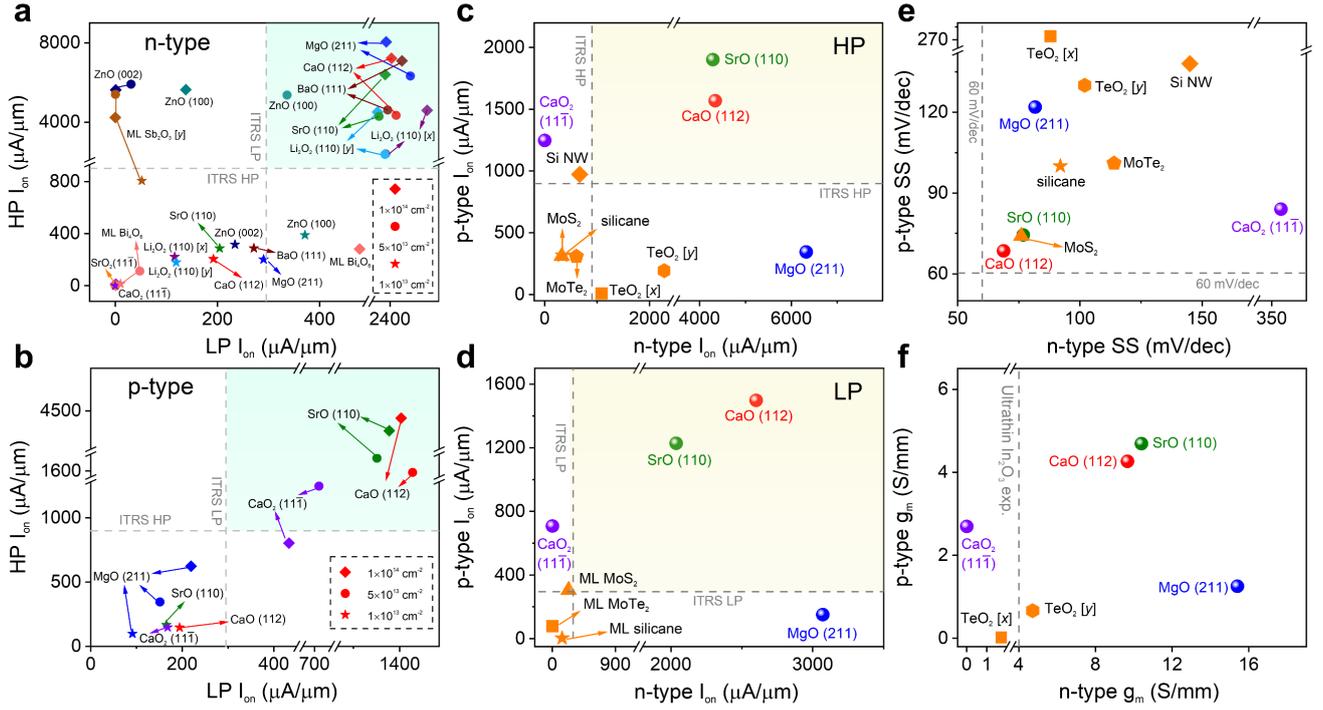

**Figure 3. Transport characteristics of the 5-nm $L_g$ FETs.** HP *vs* LP $I_{on}$ of (a) *n*-type and (b) *p*-type FETs for *Class I* and *Class II* candidates. The different doping concentrations are indicated by different markers. *P*-type *vs n*-type $I_{on}$ for the 4 oxides listed in **Figure 2(f)** under an $N_d$ = 5 × 10$^{13}$ cm$^{-2}$ for (c) HP and (d) LP applications. The simulated data of 5-nm-$L_g$ ML MoS$_2$,[37] ML MoTe$_2$,[38] ML silicane,[39] bilayer TeO$_2$ (both *x* and *y* directions),[24] and Si NW FETs[40] are also shown for comparison. In (a)-(d), the grey dashed lines represent the ITRS HP and LP criteria for $I_{on}$. (e) *P*-type *vs n*-type *SS* for the 4 oxides listed in **Figure 2(f)**, as compared with other 1D and 2D FETs. The grey dashed lines stand for the room temperature limit of *SS*. (f) Same as (e) but for transconductance $g_m$. The highest $g_m$ value of experimental ultrathin In$_2$O$_3$ FET is denoted by the grey dashed line.[41]

In addition to $I_{on}$, gate controllability is another important criterion for evaluating FET performance. Gate controllability can be described by the subthreshold swing $SS = \frac{\partial V_g}{\partial \lg I}$, and the transconductance $g_m = \frac{dI}{dV_g}$, in the subthreshold and superthreshold regions, respectively, where small *SS* and large $g_m$ are desirable. Overall, the CaO and SrO exhibit excellent *n*-type and *p*-type gate controllability in both the subthreshold and superthreshold regions (**Figures**

**3e** and **3f**). For benchmark, we also show the simulated data of ML MoS$_2$,[37] ML MoTe$_2$,[38] ML silicane,[39] bilayer TeO$_2$,[24] and Si NW[40] devices at the same $L_g$ by quantum transport simulation. The CaO and SrO possess low n-type and p-type SS of less than 80 mV/dec, which are close to the Boltzmann limit at room temperature (60 mV/dec). Moreover, these values outperform ML MoTe$_2$, ML silicane, BL TeO$_2$, and Si NW counterparts. For the superthreshold regime, the $g_m$ of CaO and SrO FETs both exceed 4 S/mm, which outperforms the bilayer TeO$_2$ FETs[24] and surpasses the highest experimentally obtained $g_m$ of n-type ultrathin In$_2$O$_3$ FET[41] (gray dashed line in **Figure 3f**). We further note that since the $I_{on}$ is determined by the field-effect mobility ($\mu_{FET}$), and $\mu_{FET}$ is proportional to $g_m$ (i.e. $\mu_{FET} = cg_m$, $c$ is a factor including the $L_g$, gate width, and gate capacitance), higher $g_m$ thus leads to larger $\mu_{FE}$ and hence higher $I_{on}$ as shown in **Figure S2**. Having good superthreshold gate controllability is thus crucial in improving the overall device performance of ultrascaled sub-1-nm thickness oxide FETs.

## 2.4 Transport mechanism

The "np-compatibility" of CaO and SrO in terms of $I_{on}$ can be better understood by examining their band structures and on-state transmission coefficients. In **Figure 4**, we use CaO, which fulfills HP/LP requirements under both p-type and n-type device configurations, and CaO$_2$(11$\overline{1}$), which fulfills HP/LP requirements only under a p-type device configuration to illustrate the underlying transport mechanisms. The electronic states around the conduction band minimum (CBM) critically influence the n-type device performance. The CBM of both CaO and CaO$_2$(11$\overline{1}$) are situated at the Γ point (**Figures 4a** and **4b**). However, the dispersion around CBM for CaO$_2$(11$\overline{1}$) is much flatter than that of the CaO, resulting in a larger electron effective mass ($m_e$) of CaO$_2$(11$\overline{1}$) (1.63 $m_0$) than CaO (0.89 $m_0$). Based on WKB approximation, the transmission probability is in a exponentially reduced by a larger carrier effective mass.[42] Thus, the n-type transmission coefficient of CaO$_2$(11$\overline{1}$) is far smaller than CaO by about 7 orders of magnitude (**Figures 4c** and **4d**), leading to the $I_{on}$ of CaO$_2$(11$\overline{1}$) (0.16 μA/μm) being 4 orders of magnitude smaller than that of CaO (4342 μA/μm).

Carrier conduction in p-type devices can be similarly explained by the hole effective mass ($m_h$) around the valence band maximum (VBM). The hole effective mass around the VBM of CaO (situated at Γ point) and of CaO$_2$(11$\overline{1}$) (situated at X point) are similar [0.91 $m_0$ for CaO and 0.937 $m_0$ for CaO$_2$(11$\overline{1}$)]. Thus, the p-type transmission coefficients for these two materials have comparable order of magnitude (**Figures 4e** and **4f**), yielding comparable $I_{on}$ in CaO (1569 µA/µm) and CaO$_2$(11$\bar{1}$) (1247 µA/µm). The slightly larger $I_{on}$ of CaO is jointly contributed by two factors: (1) $m_h$ of CaO is slightly smaller than that of CaO$_2$(11$\bar{1}$), leading to overall larger transmission coefficients in CaO; and (2) there are two degenerate bands around the VBM of CaO, compared to only one band around the VBM of CaO$_2$(11$\bar{1}$), thus enabling more channels for carrier conduction in CaO.

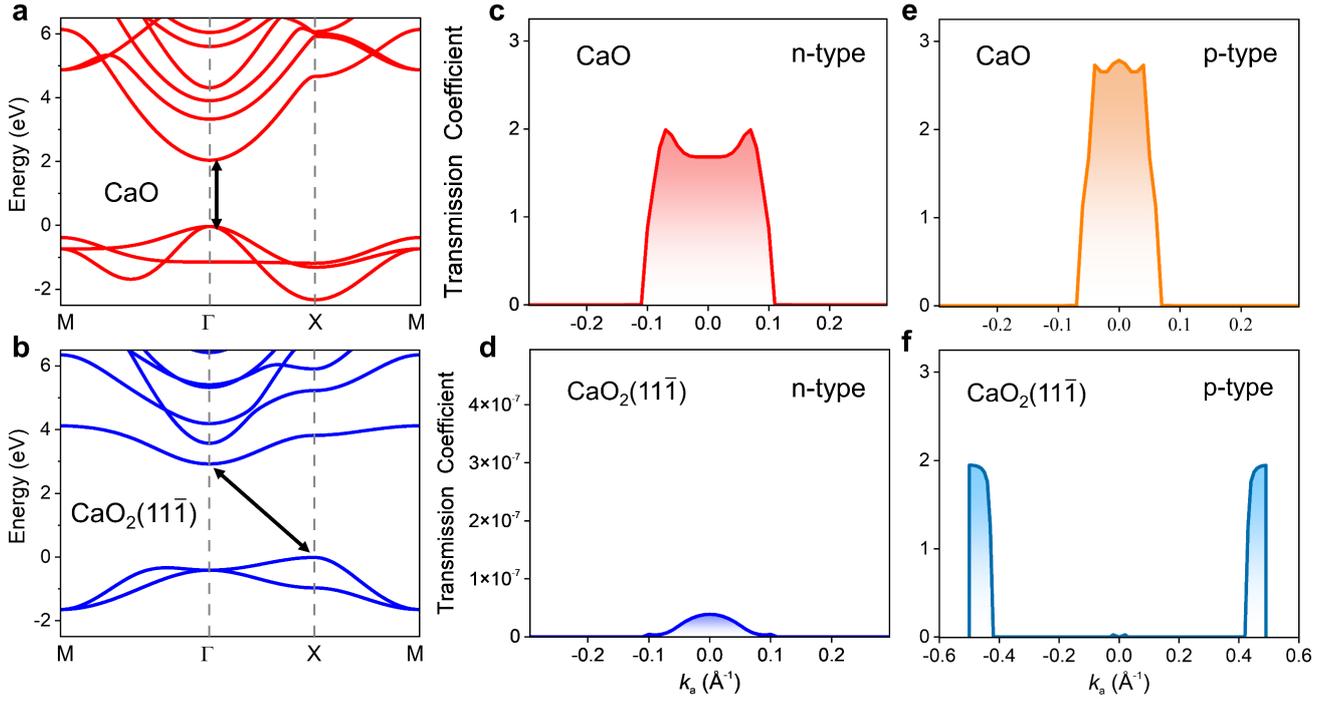

**Figure 4. Transport mechanism of CaO FET**. Band structure of (a) CaO and (b) CaO$_2$(11$\bar{1}$). The black arrow shows the gap between CBM and VBM. (c) and (d) are the transmission coefficients of *n*-type devices at the on-state for CaO and CaO$_2$(11$\bar{1}$), respectively. (e) and (f) are also the transmission coefficients but for the *p*-type devices. The transmission coefficients of CaO and CaO$_2$(11$\bar{1}$) are extracted at the same energy level.

We further note that the relatively symmetrical *n*-type and *p*-type performance in CaO FET can be explained as follows. On one hand, $m_e$ is slightly smaller than $m_h$ in CaO, which leads to the *n*-type transmission coefficient spanning over a wider *k*-vector range than that of the *p*-type counterpart (**Figures 4c** and **4e**). On the other hand, the band numbers of CaO around VBM are more than that around CBM, which leads to the higher peak value of the *p*-type device when compared to the *n*-type counterpart. The counterbalancing of these two aspects results in the *np*-symmetric behavior of CaO FET. We also perform a comparison between SrO and CaO$_2$(11$\bar{1}$) in **Section III** of **Supplementary Information** and show that the transport characteristic of SrO is akin to that of CaO.

## 2.5 Scaling performance of sub-1-nm thickness oxide transistors

We now perform a holistic assessment of the $I_{on}$, delay time ($\tau$), and power delay product (PDP) of CaO and SrO FETs when scaled below 5-nm gate length (see **Figure 5**). The simulation results of MoS$_2$, MoTe$_2$, silicane, and ultrathin In$_2$O$_3$ (only *n*-type) FETs [27, 37-39] are also included in **Figure 5** as benchmarks. For HP applications, the $I_{on}$ of both *n*-type and *p*-type devices can continue to deliver the ITRS requirements when the $L_g$ is downscaled to 2 nm and 3 nm, respectively, for CaO and SrO (**Figure 5a**). For the LP applications, the $L_g$ scaling is limited to 3 nm (4 nm) for *n*-type (*p*-type) devices for both CaO and SrO FETs (**Figure 5b**). Such $I_{on}$ scaling performances surpass MoS$_2$, MoTe$_2$, silicane, and ultrathin In$_2$O$_3$ counterparts.

The on-off switching speed and the power dissipation during the switching are characterized by the delay time $\tau$ (**Figures 5c** and **5d**) and the PDP (**Figures 5e** and **5f**), respectively, where smaller $\tau$ and lower PDP are critical to ensure faster switching speed and lower power dissipation during the on-off switching process. In terms of delay time, CaO and SrO FETs can be scaled down to (2, 3, 3, 4) nm for (*n*-type HP, *p*-type HP, *n*-type LP, *p*-type LP) devices, respectively, which outperforms MoS$_2$, MoTe$_2$, silicane, and ultrathin In$_2$O$_3$ FETs. For PDP, CaO and SrO struggle to meet the ITRS HP device requirements, with only a few examples of $L_g$ meeting the HP requirements (**Figure 5e**). In contrast, CaO and SrO perform better in terms of LP requirements, where the ITRS LP requirements can be met with $L_g$ in the ranges of 2 to 5 and 3 to 5 nm for *n*-type and *p*-type devices, respectively. The better compatibility of CaO and SrO with LP devices is consistent with the commonly observed excellent energy efficiency of ultrathin oxide semiconductor FETs.

We summarize the $L_g$ scaling limits of $I_{on}$, $\tau$ and PDP in **Figures 5g** and **5h** for CaO and SrO, respectively. The ultimate device scaling can be obtained by holistically assessing the three criteria of $I_{on}$, $\tau$ and PDP. The lowest common $L_g$ of the three criteria yields the ultimate $L_g$ scaling limit for a given device type. For CaO, the *n*-type and *p*-type devices can meet the ITRS LP requirements (**Figure 5g**) at the ultimate $L_g$ of 3 nm and 4 nm, respectively. For ITRS HP requirements, the *n*-type ultimate $L_g$ is 2 nm. However, the *p*-type device does not exhibit a common lowest $L_g$ (in the range of 2 nm to 5 nm) that can meet ITRS HP requirements. Sub-1-nm thickness CaO thus unable to deliver *np*-compatible HP device applications at the sub-5-nm $L_g$ regime. In contrast, SrO can be ultimately scaled below 5 nm for all device classes (**Figure 5h**) SrO is thus the *first-of-kind* sub-1-nm thickness oxides capable of delivering ITRS

HP and LP requirements in both n-type and p-type device configurations. Such versatile device classes of SrO are also rare among other 2D semiconductors.[24, 33] SrO ultrathin oxide could thus be promising in streamlining the fabrication process where only a single species of channel materials is needed to deliver *n*-type and *p*-type device operations for both HP and LP requirements.

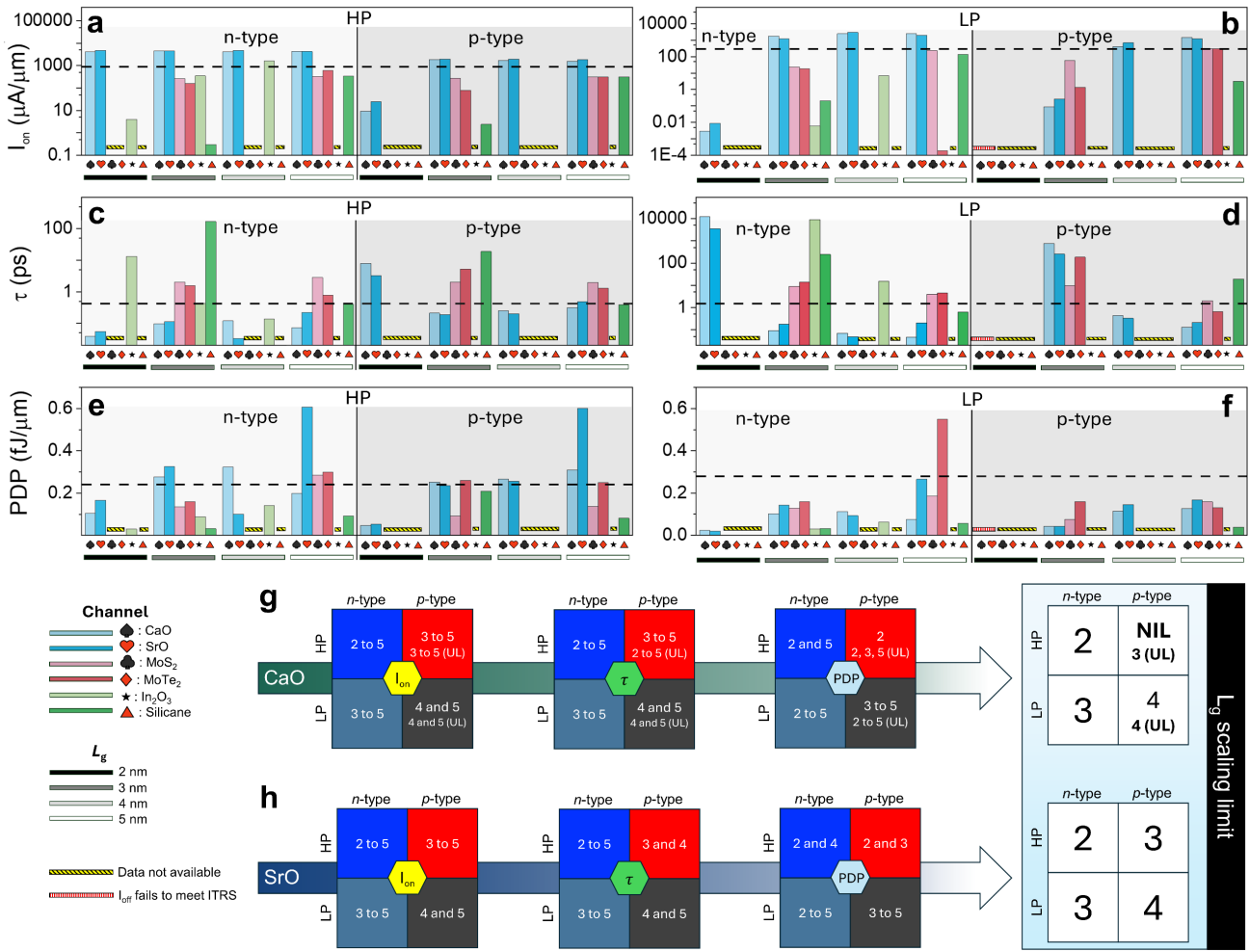

**Figure 5. Scaling performance of sub-1-nm thickness oxide transistors**. (a) HP $I_{on}$, (b) LP $I_{on}$, (c) HP $\tau$, (d) LP $\tau$, (e) HP PDP, (f) LP PDP of the CaO and SrO FETs as a function of $L_g$. The simulated data of MoS$_2$, silicane, and ultrathin In$_2$O$_3$ (only *n*-type) FETs at various $L_g$ are also shown for comparison. The black dashed lines indicate the ITRS requirements for HP and LP applications. (g) Summary of scaling performance for CaO FETs. The $L_g$ scaling limit is determined as the lowest common $L_g$ among the three device performance indicators of $I_{on}$, $\tau$ and PDP. P-type CaO FET is intrinsically incompatible with HP device operation (indicated as "NIL") as such device can only fulfill the PDP requirement at 2-nm $L_g$, which is smaller than the minimum $L_g$ of other performance indicators ($I_{on}$ and $\tau$). Adding a 1-nm UL improves the device performance, re-enabling *p*-type devices based on CaO channel fulfill HP criteria at 3-nm $L_g$. The UL in the bracket indicates the device scaling performance after 1-nm UL optimization. (h) Same as (g) but for SrO FETs. SrO FET can meet all *n*-type and *p*-type HP and LP device requirements without the need for UL optimization.

We remark that we have adopted a simplistic dual-gated FET configuration without underlap (UL) structure in the device simulations, which may be beneficial for experimental device fabrications. UL can significantly improve the performance of sub-5-nm FETs.[24-27, 33, 37-40] By including UL, we expect: (i) the scaling limit of CaO and SrO can be further pushed down; and/or (ii) more sub-1-nm thickness oxides may emerge from our screening procedures. As a proof-of-concept, we re-calculate *p*-type CaO FET with a 1-nm UL for $L_g$ = 2 to 5 nm (see **Table S8**). The device PDP, which originally could not meet the ITRS HP requirements without UL, can now meet the ITRS HP requirements at both 3-nm and 5-nm $L_g$ for *p*-type HP devices, thus re-enabling sub-1-nm-thick CaO to meet both ITRS HP and LP requirements under *p*-type device configuration (**Figure 5g**). Besides, almost all the $I_{on}$, $\tau$ and PDP for both HP and LP applications are improved with the help of 1-nm UL structure. We thus expect the device performance optimizations of sub-1-nm thickness oxides using UL and novel gate configurations such as triple gating[43] and GAA[44] to further provide a fertile ground for uncovering a wider assortment of sub-1-nm thickness oxide semiconductors for CMOS device applications (such as the 12 candidates in **Figures 2e** and **2f**) for sub-5-nm FET applications.

Previous study has predicted the bilayer $TeO_2$ as an oxide semiconductor that is simultaneously compatible with both *n*-and *p*-type HP device operations at $L_g$ = 3 nm.[24] Bilayer $TeO_2$ is absent from the ultrathin oxide screening workflow here because: (i) monolayer $TeO_2$ was eliminated at the preliminary screening stage due to their large atom number (12 atoms in a primitive unit cell); and (ii) bilayer $TeO_2$ has a thickness > 1 nm due to their bilayer morphology, which is not in alignment with the sub-1-nm channel thickness targeted in this work. It should also be noted that the *np*-compatibility of bilayer $TeO_2$ is limited to only HP operation. This behavior is in stark contrast to SrO or CaO (with 1-nm UL) which exhibits *np*-compatibility fulfilling both HP and LP criteria.

## 3. Discussion

We remark that the identifications of sub-1-nm thickness oxide semiconductors in this work shall generate a plethora of further studies for 2D oxide semiconductor devices. For instance, achieving Ohmic metal contacts[45, 46] and electrically stable dielectric interfaces[47, 48] to 2D semiconductor channels are critical device design challenges. Previous studies have demonstrated Ohmic contacts in ultrathin nonlayered $In_2O_3$ and layered $TeO_2$ FETs using Ni and $VS_2$/$NbS_2$ as the metal electrodes, respectively.[49, 50] The computational screening of Ohmic contacts to ultrathin oxides shall thus form an important basis for the experimental realization of high-performance sub-1-nm thickness oxide semiconductor devices. For gate dielectric integration, various approaches such as high-*k* layered dielectrics,[51] native oxides,[52] and metal gate electrode engineering[53] have been employed to further improve the performance of 2D semiconductor transistors. Whether such approaches can be applied to 2D oxide semiconductor FETs remains an open question. We expect the interfacial properties of sub-1-nm thickness oxides with metal and dielectrics to provide a new research frontier for future studies.

In summary, we performed an *ab initio* device-driven computational screening of ultrathin oxides with sub-1-nm thickness for ultrascaled field effect transistor application. From 3172 ultrathin layered oxides in the MC2D database and 41 nonlayered oxides, we obtained 31 preliminary oxide candidates based on first-principles calculations. Device scale length and electrical mobilty are then used to narrow down the candidate list to 12 candidates. *Ab initio* quantum transport simulations of 5-nm-$L_g$ devices of these oxide candidates revealed three previously unknown sub-1-nm-thick oxides compatible with *p*-type FET operations, namely CaO, $CaO_2$, and SrO. Notably, SrO and CaO (with 1-nm UL) are found to be first-of-kind sub-1-nm-thick oxide semiconductors that can simultaneously fulfill the ITRS HP and LP targets for both *n*-type and *p*-type devices. SrO and CaO FETs outperforms many low-dimensional scemiconductor FETs, and can be further scaled down below 5-nm gate length. Our findings expanded the pool of ultrathin oxides for future CMOS applications. The device-driven screening workflow established can be expanded for discovering other sub-classes of functional 2D semiconductors and heterosturctures for device applcations such as tunneling FET and photodetectors.

## 4. Methods

**Density functional theory calculations**

All the DFT calculations are performed in the Vienna Ab initio Simulation Package (VASP).[54] The plane wave basis set with an energy cutoff of 400 eV and projector-augmented wave (PAW) pseudopotential are adopted. For the geometry optimization, we set the force tolerance of 0.01 eV/Å on each atom and the energy criteria of $10^{-6}$ eV to obtain a reliable structure. The Monkhorst-Pack *k*-points meshes are sampled by 0.03 Å$^{-1}$ for geometry optimization and 0.02 Å$^{-1}$ for self-consistent calculation in the Brillouin zone. We employ Grimme's DFT-D3 method to include the van der Waals interaction.[55] Along the z-direction, the supercell with a vacuum space of more than 15 Å is set, and the dipole correction is used to eliminate the artificial coupling between periodic copies.[56] The GGA-PBE exchange-correlation functional is adopted for the bandgap calculation.[57]

**Dielectric constant calculations**

The static dielectric constant ($\varepsilon$) was calculated by the density-functional perturbation theory (DFPT), as implemented in VASP, including both the ionic and electronic contributions.[58] We only consider the in-plane dielectric constant ($\varepsilon_{//}$) because the transportation in the channel is along the in-plane direction. For the isotropic structure, $\varepsilon_{//}$ is the average of x and y components, namely, $\varepsilon_{//} = (\varepsilon_x + \varepsilon_y)/2$, while $\varepsilon_x$ and $\varepsilon_y$ are separated for the anisotropic structure. Since the macroscopic electric field was applied in the supercell containing the vacuum space, the contributions from the vacuum space and material itself are all included in the static dielectric tensor calculations by DFPT. Hence, we need to exclude the vacuum contribution based on the following formula:[59]

$$\varepsilon_{//}^m = 1 + \frac{L}{t}(\varepsilon_{//}^{sup} - 1) \qquad (2)$$

where $\varepsilon_{//}^{sup}$ is the supercell dielectric constant, $\varepsilon_{//}^m$ is the material dielectric constant, L is the supercell height, and t is the thickness of the monolayer structure. t is obtained from the interlayer distance of the corresponding bilayer.[59]

**Mobility calculations**

The carrier mobility $\mu$ is computed based on the deformation potential theory (DPT).[60] For the anisotropic structure, $\mu$ along the transport direction (assumed as x direction) is obtained by the following formula: [61]

$$\mu_x = \frac{e\hbar^3(\frac{5C_{2D,x} + 3C_{2D,y}}{8})}{k_B T m_x m_d (\frac{9E_{1,x}^2 + 7E_{1,x}E_{1,y} + 4E_{1,y}^2}{20})} \qquad (3)$$

where $\hbar$, e, $k_B$, and T represent the Planck constant, electron charge, Boltzmann constant, and temperature (set to 300 K), respectively. $m_x$ is the x-direction effective mass, and $m_d$ is the average effective mass between the x and y directions, as calculated by $m_d = \sqrt{m_x m_y}$. $E_1$ stands for the deformation potential, which can be computed by $E_1 = \Delta E/(\Delta l/l_0)$. Here $l_0$ is the lattice parameter, $\Delta l$ is the variation of $l_0$ with a step of 0.5% from -2% to 2%, and $\Delta E$ is the energy variation of the band edge, namely, conduction band minimum variation for electron and valence band maximum variation for hole. The elastic modulus $C_{2D}$ is defined as $C_{2D} = 2[\partial^2 E/\partial(\Delta l/l_0)^2]/S_0$, where $S_0$ and $E$ are the lattice area ($x$ and $y$ directions) and the total energy after optimization, respectively. For the isotropic structure, $C_{2D,x}$ and $E_{1,x}$ have the same values as $C_{2D,y}$ and $E_{1,y}$. Therefore, equation (2) becomes:

$$\mu_x = \mu_y = \frac{e\hbar^3 C_{2D}}{k_B T m_x m_d E_1^2} \quad (4)$$

**Device Simulations**

The device transport characteristics are simulated by combining DFT and non-equilibrium Green's function (NEGF). Multiple DFT-NEGF quantum device simulation codes exist,[34, 62] and we have chosen the DFT-NEGF method as applied in the QuantumATK 2023 for this work.[33, 63] In an FET, there are three major components, the source electrode, drain electrode, and channel region. The interaction between channel and source/drain electrodes is described by the self-energy $\Sigma_{k_{//}}^{l/r}$, where $k_{//}$ is the surface-parallel reciprocal lattice vector and $l/r$ represents the left (source)/right (drain) electrodes. Based on $\Sigma_{k_{//}}^{l/r}$, we can obtain the broadening matrix $\Gamma_{k_{//}}^{l/r}(E) = i[\Sigma_{k_{//}}^{l/r} - (\Sigma_{k_{//}}^{l/r})^\dagger]$ and the retarded [advanced] Green's function $G_{k_{//}}(E)$ [$G_{k_{//}}^\dagger(E)$]. Thus, the transmission coefficient $T_{k_{//}}(E)$ can be calculated by the following formula:

$$T_{k_{//}}(E) = Tr[\Gamma_{k_{//}}^l(E) G_{k_{//}}(E) \Gamma_{k_{//}}^r(E) G_{k_{//}}^\dagger(E)] \quad (5)$$

To obtain the transmission function $T(E)$, we average the $T_{k_{//}}(E)$ over $k_{//}$ in the irreducible Brillouin zone. The drain current ($I_{ds}$) is given by the Landauer–Büttiker formula:

$$I_{ds} = \frac{2e}{h} \int_{-\infty}^{+\infty} [f_D(E - \mu_D) - f_S(E - \mu_S)] T(E) \, dE \quad (6)$$

where $f_D$ ($f_S$), and $\mu_D$ ($\mu_S$) denote the Fermi-Dirac distribution function of drain (source) electrode, and the electrochemical potential of drain (source) electrode, respectively. In our simulation, we adopt the PseudoDojo pseudopotential and set the temperature as 300 K. The $k$-point meshes are sampled by 8×1×270 for all the devices. Along the transverse, vertical, and transport directions, we employ the Periodic, Neumann, and Dirichlet boundary conditions, respectively.

The source and drain electrodes are doped by the electron ($n$-type) or hole ($p$-type) based on the atomic compensation charge method (see **section IV** in supplementary information for details). The doping concentration $N_d$ can significantly influence the device performance. In the focus studies of CaO and SrO FET performance, a contact doping level of $N_d = 5 \times 10^{13}$ cm$^{-2}$ is used. Such doping level corresponds to $1{\sim}2 \times 10^{20}$ cm$^{-3}$ in the 3D case, which is comparable to the experimental $N_d \sim 9 \times 10^{19}$ cm$^{-3}$ of ultrathin In$_2$O$_3$ ($t$ = 3.5 nm),[17] thus indicating the feasibility of achieving such $N_d$ experimentally. In addition to the current-voltage characteristics, we also calculate the delay time ($\tau$) and the power delay product (PDP). Delay time is defined as $\tau = C_t V_{dd}/I_{on}$, where $C_t$ and $V_{dd}$ are the total capacitance and supply voltage, respectively. Based on the ITRS criteria, $C_t = 3\partial Q_{ch}/\partial V_g$ where $Q_{ch}$ indicates the total charge in the channel. PDP can be calculated as PDP = $V_{dd} I_{on} \tau = C_t V_{dd}^2$. The viability of the DFT-NEGF method for sub-10-nm device simulation has been demonstrated in the 5-nm-$L_g$ carbon nanotube FETs, which shows good agreement in the transport characteristics between simulation and experiment.[64]

We note that the PBE functional will underestimate the bandgap of semiconductors.[65] However, in a device setup, the electron-electron coupling of the channel material can be screened by the dielectric environment as well as the doping carriers from the electrodes, resulting in the depression of many-body effects and thus the accurate estimation of PBE bandgap. For example, previous studies show that the bandgap of ML MoS$_2$ sandwiched by high-$\kappa$ dielectrics at the GW level is decreased from 2.8 to 1.9 eV,[66] in agreement with that at the GGA-PBE level (1.76 eV).[67] On the other hand, the GW bandgap of intrinsic ML MoSe$_2$ is renormalized to 1.59 eV at a degenerately doped state,[68] which is consistent with the GAA (1.52 eV) bandgap.[69] Since our focus is on the transport properties of ultrathin oxides as the channel materials, we expect the PBE functional to be sufficient when evaluating the bandgap values of the candidate oxides.

**Data Availability Statement**

The data that support the findings of this study are available from the corresponding authors upon reasonable request.

**Conflict of Interest**

The authors declare no conflict of interest.

**Acknowledgment**

This work is supported by the Ministry of Science and Technology of China (No.2022YFA1203904 and No. 2022YFA1200072), the National Natural Science Foundation of China (No. 91964101, No. 12274002, and No. 12164036, No. 62174074), the Fundamental Research Funds for the Central Universities, the Natural Science Foundation of Ningxia of China (No. 2020AAC03271), the youth talent training project of Ningxia of China (2016), and the High-performance Computing Platform of Peking University. Linqiang Xu and Y.S.A. acknowledge the support of Singapore University of Technology and Design Kickstarter Initiatives (SKI) under the Award No. SKI 2021_01_12. Y.S.A. is also supported by the SMU-SUTD Joint Grant (Award No. 22-SIS-SMU-054) and SUTD-ZJU Thematic Research Grant (Award No. SUTD-ZJU (TR) 202203). The computational work for this article was partially performed on resources of the National Supercomputing Centre, Singapore (https://www.nscc.sg). C.S. Lau acknowledges support by the Agency for Science, Technology, and Research (A*STAR) under its MTC YIRG grant No. M21K3c0124 and MTC IRG grant No. M23M6c0103.

**Authors Contributions**

Y.S.A. and J. L. designed and supervised the project. Linqiang Xu (L.X.) performed ab initio DFT calculations and NEGF device simulations. Y.H. provided the DFT simulation data of nonlayered oxides. Y.S.A. and L.X. performed data visualizations. J.L. and Lianqiang Xu provided part of the computing resources for NEGF device simulations. L.X., Y.S.A. and L. J. analyzed the data and wrote the manuscript with inputs from all other authors. The manuscript reflects the contributions of all authors.19
**Acknowledgment**

This work is supported by the Ministry of Science and Technology of China (No.2022YFA1203904 and No. 2022YFA1200072), the National Natural Science Foundation of China (No. 91964101, No. 12274002, and No. 12164036, No. 62174074), the Fundamental Research Funds for the Central Universities, the Natural Science Foundation of Ningxia of China (No. 2020AAC03271), the youth talent training project of Ningxia of China (2016), and the High-performance Computing Platform of Peking University. Linqiang Xu and Y.S.A. acknowledge the support of Singapore University of Technology and Design Kickstarter Initiatives (SKI) under the Award No. SKI 2021_01_12. Y.S.A. is also supported by the SMU-SUTD Joint Grant (Award No. 22-SIS-SMU-054) and SUTD-ZJU Thematic Research Grant (Award No. SUTD-ZJU (TR) 202203). The computational work for this article was partially performed on resources of the National Supercomputing Centre, Singapore (https://www.nscc.sg). C.S. Lau acknowledges support by the Agency for Science, Technology, and Research (A*STAR) under its MTC YIRG grant No. M21K3c0124 and MTC IRG grant No. M23M6c0103.

**Authors Contributions**

Y.S.A. and J. L. designed and supervised the project. Linqiang Xu (L.X.) performed ab initio DFT calculations and NEGF device simulations. Y.H. provided the DFT simulation data of nonlayered oxides. Y.S.A. and L.X. performed data visualizations. J.L. and Lianqiang Xu provided part of the computing resources for NEGF device simulations. L.X., Y.S.A. and L. J. analyzed the data and wrote the manuscript with inputs from all other authors. The manuscript reflects the contributions of all authors.

**Table 1.** Benchmark of the CaO and SrO FETs against the IRDS 2023 requirements at various technology nodes for the HP and HD transistors.

| Oxide Formula | Carrier Type | $I_{on}^{HP}$ (µA/µm) | $I_{on}^{HD}$ (µA/µm) |
|---|---|---|---|
| CaO | n | 3856 | 3037 |
|  | p | 1777 | 1540 |
| SrO | n | 3953 | 2775 |
|  | p | 1735 | 1406 |
| IRDS 2023 Requirements | Technology Nodes (nm) | $I_{on}^{HP}$ (µA/µm) | $I_{on}^{HD}$ (µA/µm) |
|  | 3 | 874 | 644 |
|  | 2 | 787 | 602 |
|  | 1.5 | 759 | 546 |
|  | 1 | 775 | 562 |
|  | 0.7 | 771 | 570 |
|  | 0.5 | 790 | 587 |